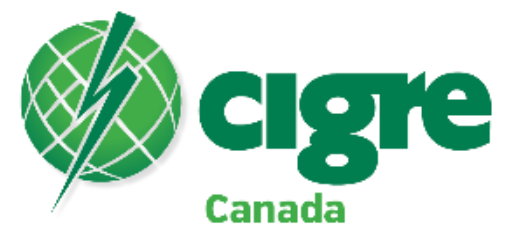

10341
C4 POWER SYSTEM TECHNICAL PERFORMANCE
PS1 System Enhancement, Markets and Regulation

# Stability Enhancement in Weak Grids with High Renewable Penetration: Synchronous Condenser vs. Converter-based Technologies

**Rasool Heydari***
**Hitachi**
**Sweden**
Rasool.heydari@hitachienergy.com

**Prabhat Ranjan Bana**
**Hitachi**
**Sweden**
prabhat.ranjan-bana@hitachienergy.com

**Jean Philippe Hasler**
**Hitachi**
**Sweden**
jean-philippe.hasler@hitachienergy.com

**Anders Bostrom**
**Hitachi**
**US**
anders.x.bostrom@hitachienergy.com

**Mikael Halonen**
**Hitachi**
**Sweden**
mikael.halonen@hitachienergy.com

**SUMMARY**

Integrating renewable energy sources into weak grids presents significant challenges for maintaining grid stability. Several essential services are required to ensure a stable and secure grid, including voltage support, fault current injection, system strength support, frequency control, and inertia support. Synchronous condensers, with their inherent inertia and fault current capabilities, have been traditional solutions for grid stability. Recent advancements in power electronics and control structures have led to the development of various technologies aimed at enhancing power grid stability. Static synchronous compensators (STATCOM) and Enhanced STATCOM, which can provide active and reactive power support, show considerable promise. These technologies offer grid-forming (GFM) capabilities that can significantly improve the stability of weak grids with high renewable penetration. Advancements in STATCOM and E-STATCOM technologies offer enhanced control tunability, controllable damping, and immediate response to grid perturbations. This paper investigates the comparative performance of synchronous condensers versus STATCOM and Enhanced STATCOM with grid-forming capability in providing services and enhancing the stability of weak grids with high renewable penetration.

# 1 Introduction

The integration of renewable energy sources (RES) into electrical grids is a crucial step towards achieving sustainable energy systems. However, this transition introduces significant challenges, particularly for weak grids, which are more susceptible to instability [1, 2]. Ensuring grid stability in such environments necessitates a suite of essential services, including voltage support, fault current injection, system strength support, frequency control, and inertia support. Recent advancements in power electronics and control systems have led to the development of various technologies aimed at enhancing grid stability. Converter-based technologies such as Static Synchronous Compensators (STATCOM) and Enhanced STATCOM (E-STATCOM) equipped with active power support have shown considerable promise. These technologies represent a significant leap forward in addressing the stability issues posed by high renewable penetration [3, 4].

Conventionally, power converters have employed grid-following control (GFL), which adjusts their output to align with the grid's voltage and frequency. However, the challenges associated with independently controlling grid voltage and frequency have led to the development of grid-forming control (GFM). GFM technology enables STATCOMs to shape the grid voltage actively, thereby reducing their reliance on the grid's existing voltage and frequency conditions [5, 6]. As early as 2025, the global footprint of GFM STATCOMs was surpassing 10 000 MVAr and today there are >20 installations in service for STATCOM with GFM capability [10]. Several E-STATCOM projects are also under execution.

Synchronous condensers, with their inherent inertia and fault current capabilities, have been traditional solutions for grid stability. However, the performance of synchronous condensers is limited by the machine's physical characteristics. In contrast, advancements in STATCOM and E-STATCOM technologies, with GFM capabilities, offer enhanced control tunability and immediate response to grid perturbations. E-STATCOMs emulate the inertial response of traditional generators, providing synthetic inertia and robust frequency control. These systems can dynamically adjust their parameters to optimize grid support, making them highly adaptable to varying grid conditions and more effective in supporting weak grids with high renewable penetration.

This paper aims to investigate the comparative performance of synchronous condensers versus STATCOM and E-STATCOM technologies in enhancing the stability of weak grids with high renewable penetration. Through a detailed analysis of their capabilities in providing essential grid services, this study contributes to understanding how these technologies can be leveraged to ensure a stable and secure grid.

# 2 Essential Services for Grid Stability

Ensuring stability in power grids, particularly in weak grids with high penetration of RES and low short-circuit levels, necessitates a comprehensive suite of essential services. These services are critical to maintaining the reliability and resilience of the grid, especially as the integration of RES continues to grow. This section elaborates on the key services required to achieve grid stability under these challenging conditions.

## 2.1 Grid Strength

The Grid Strength refers to the robustness and stability of an electrical power grid. It indicates how well the grid can handle disturbances, such as fluctuations in power demand, supply variations, or faults. The Grid Strength defines the Short-Circuit Strength (Fault Level), an assumption that is valid for a traditional grid equipped with a synchronous machine but not necessarily for a grid populated with Renewable Energy Sources (RES), which may provide fault current without necessarily contributing to the grid's strength.

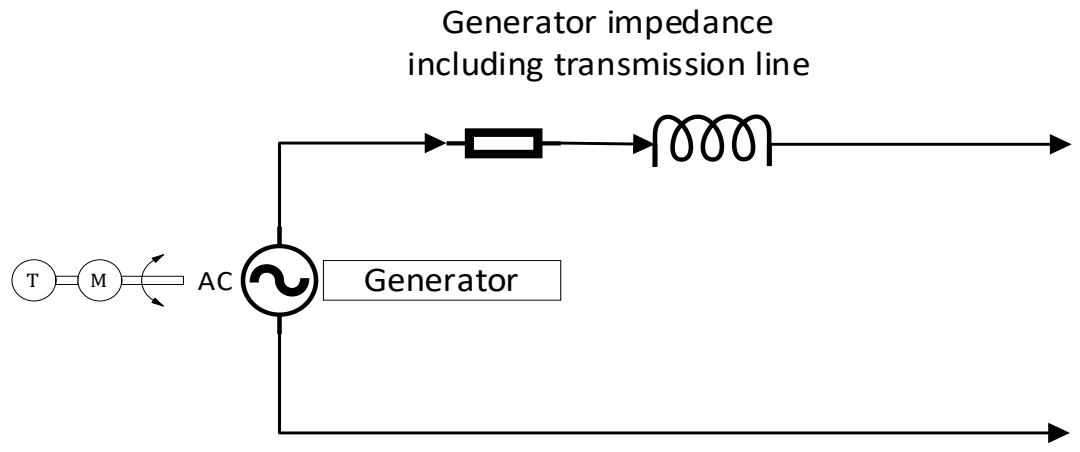

*Figure 1 – Impedance modelling of a synchronous machine.*

A voltage source and an impedance can be used to synthesize a synchronous machine, as shown in Fig. 1. The grid strength is obtained from the instantaneous response from the stiff voltage source and constant impedance. After the initial support, the support from the voltage source is reduced (in terms of phase angle) due to the available energy stored in the rotating mass, providing limited inertia support. For extreme voltage dip (fault), the initial current is reduced due to the non-constant impedance. The synchronous machine participates in the fault current, defined by its total impedance, magnitude, and time response, which typically occurs within a few milliseconds. The rotating masses, constituted by the machine rotor and of at least one coupled turbine or flywheel, create power oscillations at around 1 Hz (total mass against the power system), namely called Power Oscillation requiring a Power system stabilizer (PSS) to mitigate the damping, and at a higher frequency between 10 Hz up to 45 Hz, oscillation caused by the multi-mass and mass coupler causing so-called Sub-Synchronous Resonance (SSR), which may require damping in the electrical network [7, 8].

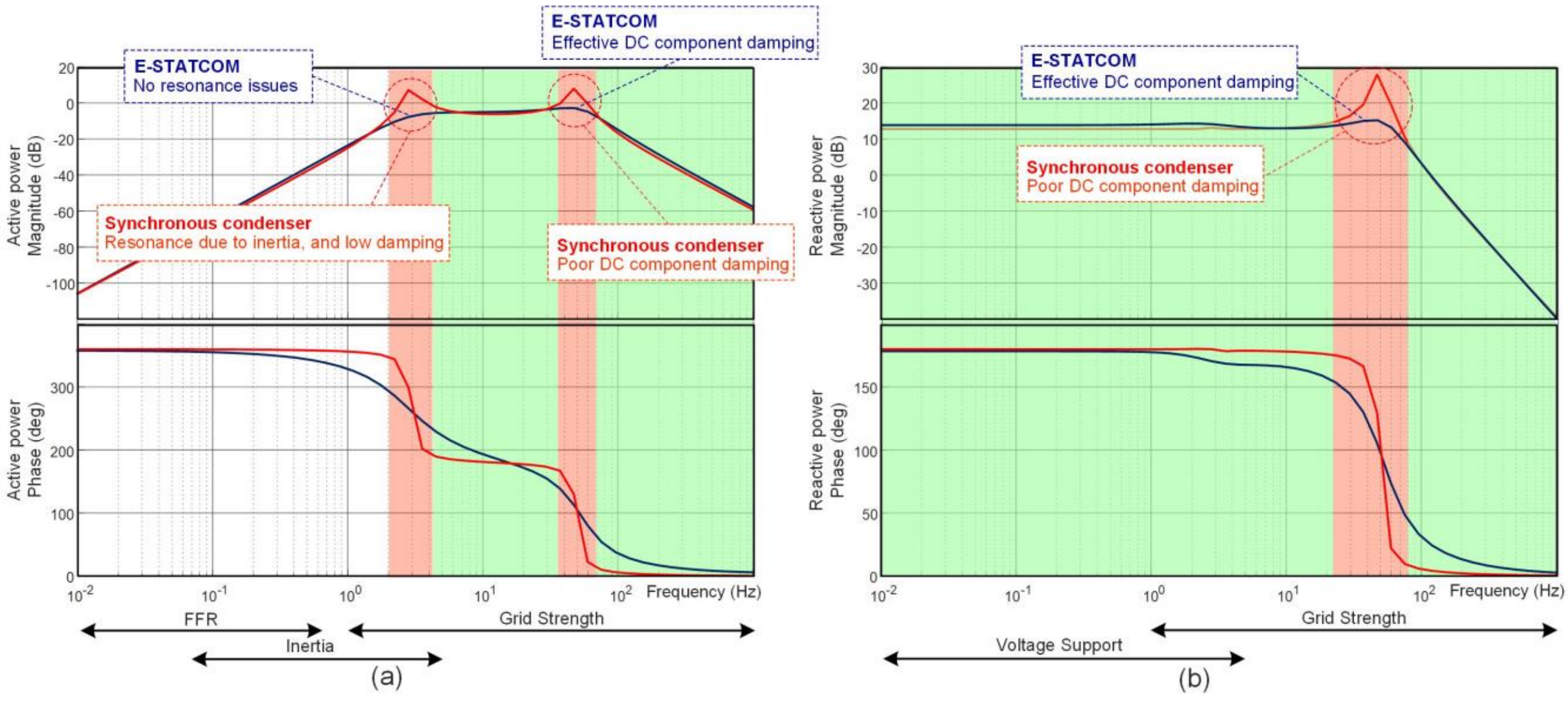

*Figure 2 – a) Active power response to phase angle modulation. b) Reactive power response to voltage amplitude perturbation.*

Fig. 2 (a) illustrates a comparative analysis of the active power response of the GFM E-STATCOM and a synchronous condenser with an equal inertia constant to phase angle modulation at the grid. This plot visually represents the amplitude and phase of the active power response of the devices when the grid voltage waveform phase angle is modulated with sub-synchronous frequencies. This technique provides a graphical insight into potential interactions (e.g., unstable oscillatory modes). For example, the synchronous machine exhibits SSR at around 2-3 Hz. However, the GFM STATCOM is designed to smoothly provide power damping for the SSR and avoid resonances at low frequencies. Fig. 2 (b) shows the reactive power response of the GFM E-STATCOM and a synchronous condenser to voltage amplitude perturbation. The poor DC component damping of the synchronous condenser is also evident from both Fig. 2 (a) and (b), with a resonance at f=50 Hz.

Traditional Inverters, GFL-STATCOM, Classic SVC, and Classic HVDC converters do not instantaneously respond to disturbances, as they all behave as current sources controlled by Power Control and Reactive Power Control (or voltage control) based on a closed-loop control, i.e., delaying the grid support. This converter may participate in fault current by using control actions that delay the response by several tens of milliseconds. A converter equipped with a harmonic filter or TSC, characterizing the converter impedance as a capacitor, reduces the grid strength. A GFM converter can be represented as an inductance, series resistances, and a voltage source, with the series resistance accounting for damping, where its energy is stored or supplied by the converter's energy storage. The converter's response to grid perturbations is immediate (<50 us) thanks to the open-loop control that emulates the virtual inductance and resistance. This response defines the grid-forming capability as a synchronous machine with a selected inductance, damping (represented by resistance), and inertia. The power output required to support or minimize the grid phase angle deviation must be limited by the converter's energy availability. The required energy to maintain transient stability depends on the other connected converters that restore power balance in the grid.

At a minimum, a GFM STATCOM must have a minimum energy (stored in the MMC DC capacitor), which controls its power through power control, with a response time of >20 ms. This ensures the damping of sub-synchronous oscillations introduced by power electronic inverters controlling their power or reactive power output through a closed-loop controller. For loss of power generation, the GFM converter must provide a significantly greater amount of energy until the fast frequency response service restores the power balance, which can occur within a few seconds. The energy support may be in both directions: positive if a power unit experiences a temporary or permanent power drop and negative if an excess power generation needs to be absorbed. For the reduction of the transmission path resulting in a reduction of grid strength at a particular node, the GFM STATCOM must prevent overloading of the congested line by, for example, absorbing the excess power generated by power units utilizing the congested transmission path to deliver the power.

## 2.2 DC component mitigation

Synchronous condensers, as well as transformers, reactors, and other inductive equipment, inherently exhibit a high direct current (DC) component during faults. This DC current adversely affects the system and is characterized by resonance at the fundamental frequency, as illustrated in Fig. 2. This resonance can lead to challenges for circuit breakers to open and interrupt fault current, referred to as delayed current zero crossing or zero-miss. The concerns with this are breaker wear and tear, eventual failure, and impaired system selectivity. In addition, a high DC offset may create issues with converter control when grid losses or the resistive damping between units is very low. In cases where requirements restrict using the virtual impedance to tune damping, sharp resonances at power frequency can risk the instability of a grid-forming converter with very low losses. To address this, the GFM STATCOM is designed with an impedance that effectively damps these DC components by incorporating fictitious losses or virtual resistance.

In conclusion, synchronous condensers exhibit high DC components during faults, affecting system stability and posing challenges for circuit breakers. The GFM STATCOM effectively damps these DC components, preventing issues like delayed current zero crossing and reducing the risk of instability in low-loss grids.

## 2.3 Inertia and Frequency Control

Inertia support is a fundamental aspect of maintaining frequency stability in the power grid. The inertia support provides active power, generating power during frequency drops and absorbing power during frequency increases. The initial amount of power is defined by the sub-transient reactance, which ensures stable power injection or absorption by allowing it to increase, if necessary, the grid strength. Traditionally, inertia refers to the energy stored in large rotating generators and specific industrial motors, which helps to resist changes in frequency by providing a temporary buffer during disturbances. This stored energy is crucial for the initial response to frequency deviations, allowing time for other control mechanisms to activate.

As the grid increasingly incorporates inverter-based renewable energy sources, which do not inherently provide inertia, the role of synthetic inertia becomes more significant. Synthetic inertia, provided by advanced power electronics and control systems, can emulate the inertial response of traditional generators, thereby supporting frequency stability. Adequate inertia support, in conjunction with frequency control measures, ensures that the power grid can maintain equilibrium between load and generation, preventing instability and potential blackouts. Frequency stability in the power grid is paramount for maintaining the equilibrium between load and generation, ensuring the system operates within its designated frequency range. This balance is crucial, as deviations can lead to system instability and potential power outages. Frequency control is required across various timescales, ranging from sub-second responses to adjustments over several minutes. Key aspects of frequency stability, shown in Fig. 3, include the frequency nadir, which is the lowest point of frequency reached following a disturbance, and the Rate of Change of Frequency (RoCoF), which measures how quickly the frequency changes. The frequency nadir is a crucial indicator of the system's ability to withstand disturbances, while RoCoF is essential for detecting and responding to rapid frequency deviations. Effective management of these parameters is vital for maintaining grid stability and preventing cascading failures.

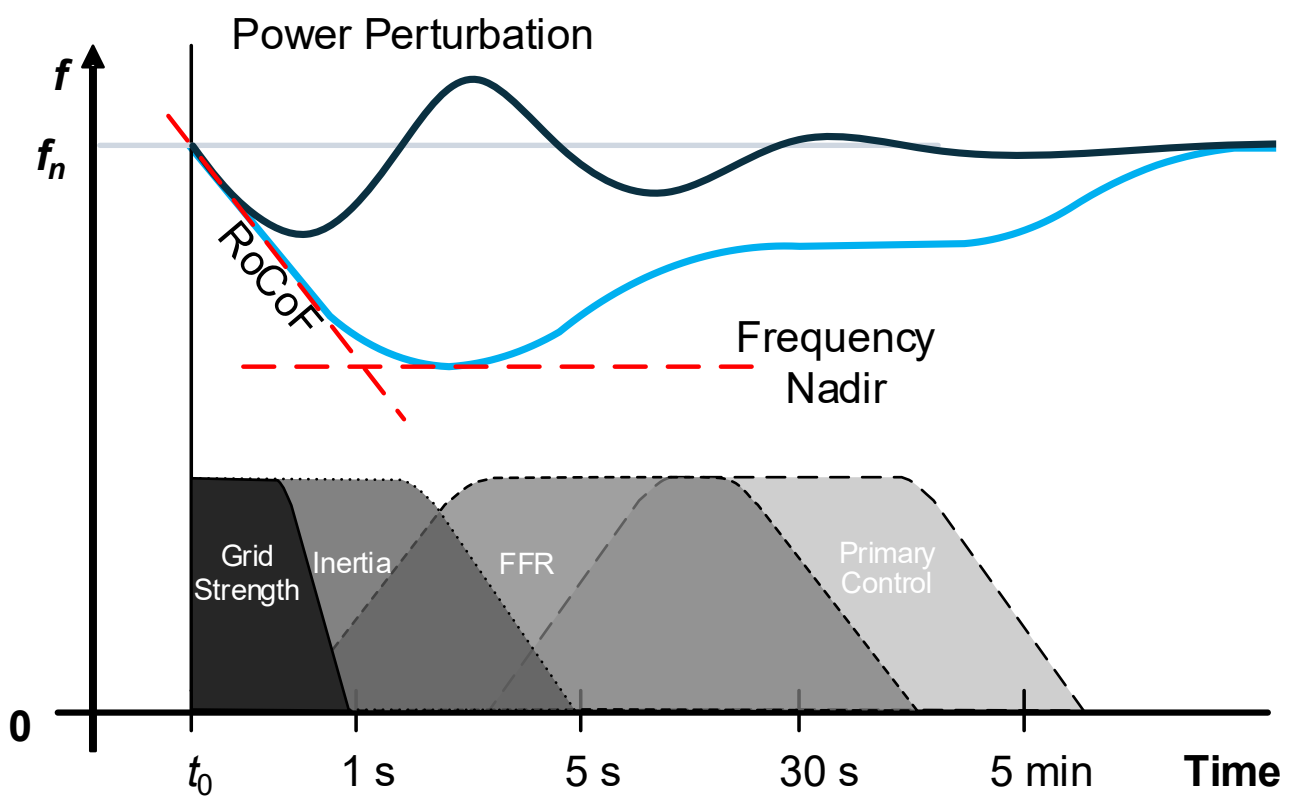

*Figure 3* – ***Power perturbation and frequency response***

### 2.4 Energy utilization

From an inertia perspective, Synchronous Condensers and E-STATCOMs can provide support, as discussed in Section 2.3. The synchronous Condenser inertia constant H is quantified by the measure of the machine kinetic energy stored in its rotating mass, normalized by the nameplate rating. The inertia is captive to machine design and dispensed in proportion to the connecting system power frequency drop, counteracting the change in speed. In this context, the behavior of the Synchronous Condenser helps to limit the RoCoF to decline gradient and nadir, giving reliability-must-run generation time to respond and reaccelerate the system by injecting active power. During this process, as equilibrium is reached between system frequency and Synchronous Condenser rotational speed, the machine to system energy exchange gradually reduces and eventually reverses to reaccelerate the machine's nominal speed as the system frequency recovers to its nominal value. During this process, the energy exchange between the machine and the power system is directly proportional to the frequency change, $\Delta E \approx 2 \times \Delta\omega \times H$. This process, being entirely governed by the laws of physics and machine mechanical properties, is not adjustable or controllable for a given machine in active operation. This means that even for a large machine with a high H-factor, the available energy is limited, i.e. only around 5% to 10% of the machine rating is used for energy exchange. Considering that the Synchronous Condenser must operate in synchronism with the power system, there are limitations concerning frequency variations. Depending on frequency excursion severity, the Synchronous condenser may risk tripping due to under or over-frequency to avoid dropping out of synchronism.

In general, and for the weaker system in particular, the E-STATCOM presents the better alternative. Like the Synchronous Condenser, in response to frequency excursions, the E-STATCOM can inject and absorb active power as required by the system. From a power system perspective, both devices appear to show similar behavior. This is evident by the E-STATCOM Grid Forming control, which makes it behave like a Virtual Synchronous Machine (VSM). The difference between the Synchronous Condenser and the E-STATCOM from a performance perspective is the controllability and the magnitude of inertia support provided by the respective devices. The E-STATCOM is fully controllable and able to inject close to its maximum charged power during a RoCoF event. Its regulator is tunable to an extent that can avoid oscillations typically seen for machines. In conclusion, for inertia support, the E-STATCOM can efficiently utilize all its stored energy, whereas the Synchronous Condenser only provides a fraction of its inherent kinetic energy during RoCoF events related to the variation in frequency. The ability to tune and condition the E-STATCOM active and reactive support makes it extremely efficient in utilizing its inherent energy.

## 2.5 Voltage control

The E-STATCOM and STATCOM, through their grid forming behavior provide stable and immediate voltage response also for conditions where the device rating is on par with or higher than the connecting system fault level. Compared with Synchronous condensers, STATCOM devices are symmetrical and scalable in rating, whereas Synchronous condensers are captive to machine parameters, excitation properties and related physical constraints, giving the machine an asymmetrical reactive dynamic range resulting in roughly 50% inductive range based on nameplate rating.

The voltage source behind the impedance behavior of the grid forming control makes the dynamic response to grid side voltage changes immediate. Slower voltage variations due to daily or longer-term load changes are handled by a conventional AC Voltage regulator. This voltage regulator can be seen as setting the "back emf" of the grid forming voltage source and handling the reference voltage. One of the benefits of this separation of voltage scheduling and dynamic response is that the AC voltage regulator can operate at a very low gain, providing a slow response to steady state variations. This eliminates the need for gain control provided by gain reduction and gain optimization control functions often seen in grid following converters. Seen from the high voltage terminals, the STATCOM or E-STATCOM has positive damping throughout the entire frequency range, staying passive or having a damping effect on external harmonics and resonances. The inherent frequency characteristics make these devices ideal for integration of Power Oscillation Damping (POD) and Power System Stabilizer (PSS) control functions.

## 2.6 Fault Current Capability

Synchronous condensers are known for their high fault current capability. It's fault current level is defined by its sub-transient reactance, governed by machine parameters combined with the step-up transformer impedance. Its terminal fault current is characterized by a high magnitude DC component and relatively poorly damped oscillation. The fault current magnitude is typically 3-4 per unit, DC offset excluded.

On the other hand, for a GFM STATCOM to effectively support full fault current capability, it must exhibit symmetric impedance for both the positive and negative sequence components. This means that during asymmetrical faults, the fault current should be proportional to the voltage dip in both the positive and negative sequences. To meet this requirement, the GFM STATCOM should utilize either a 2-level or 3-level converter type or a Modular Multilevel Converter (MMC) configured in a double-wye arrangement. Further, the GFM STATCOM allows for controllable fault current with a selectable DC component damping of 1 to 3 cycles. The converter impedance can be adjusted similarly to that of a synchronous machine, including its transformer, though the maximum overload capability of the valve current may limit it.

Using an overcurrent-rated semiconductor (designed for short-time overload) may restrict the effectiveness of the fault current limiter.

In conclusion, while Synchronous condensers can contribute to higher fault current injection, they exhibit limited control over DC components during fault intervals. In contrast, GFM STATCOMs and E-STATCOMs effectively manage DC components during faults. Additionally, they can contribute similar fault current like Synchronous condensers with an appropriate rating. However, the fault current contribution is limited to the converter rating.

# 3 Simulation in the Loop (SIL) Results

A SIL analysis was conducted in an EMT simulation environment to validate the discussed theory. Fig. 4 shows the power system network with constant GFL STATCOM and PE loads. The wind farm connects to the 400kV grid via long-cables, leading to issues such as DC presence, transient overvoltage (TOV), and SSO during operational changes or transient events. Therefore, a compensating device like Synchronous Condensers or GFM E-STATCOM is needed to stabilize the grid. The following analysis compares their performance under various transient conditions, with both devices having the same inertia constant (2s) and sub-transient reactance (20%).

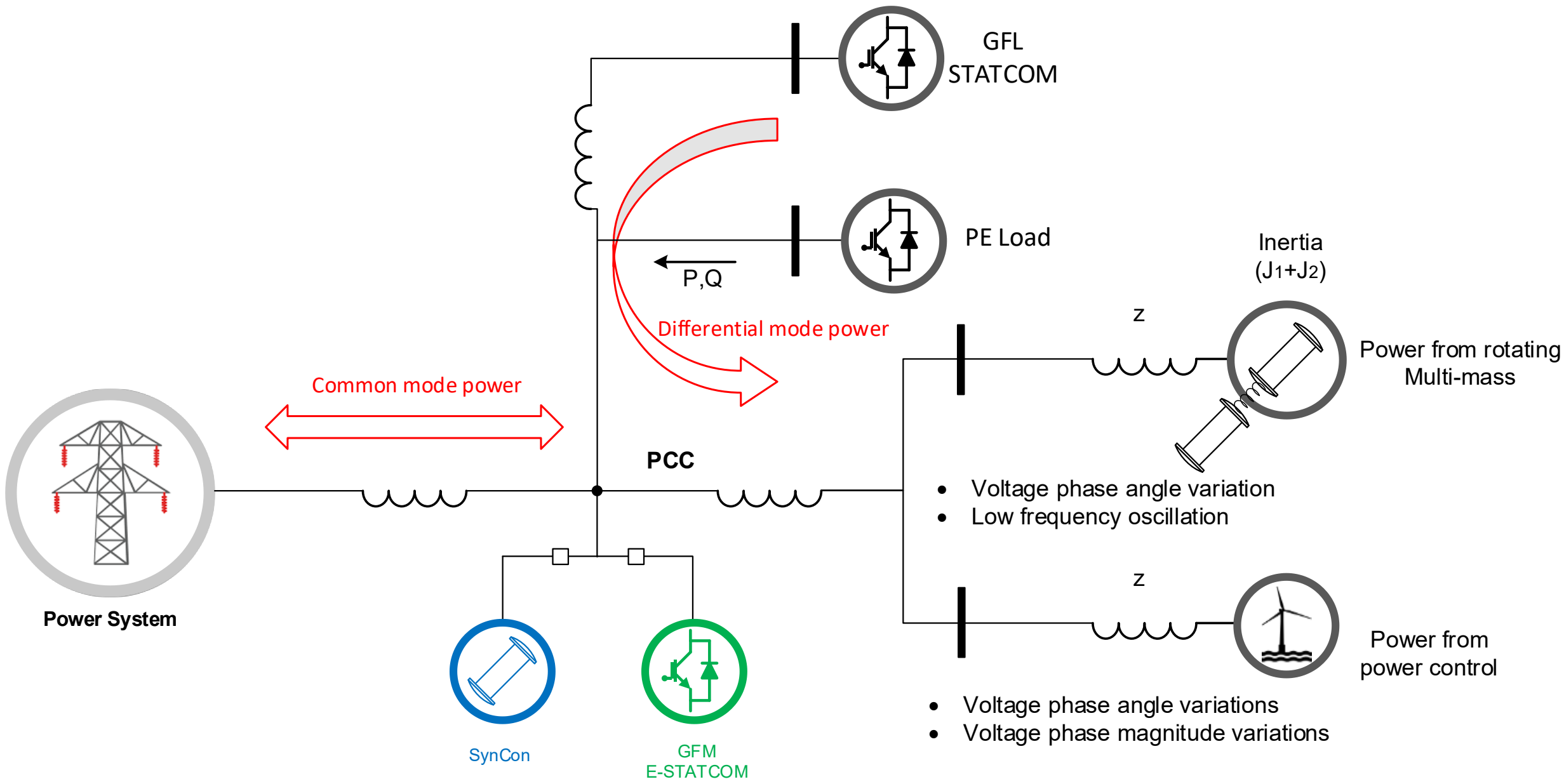

*Figure 4 – Schematic of the power system network for SIL analysis*

## 3.1 DC component

Phase jumps between 10° and 30° commonly occur in todays power systems with large integration of renewables, requiring compensating devices to dampen the resulting DC. Fig. 5 shows simulation results for a +30° phase jump event. The Synchronous Condensers take longer to damp the DC from the primary side voltage ($u_{Pri}$) and power injected ($pq_{Pri}$), while the E-STATCOM quickly and effectively mitigates it.

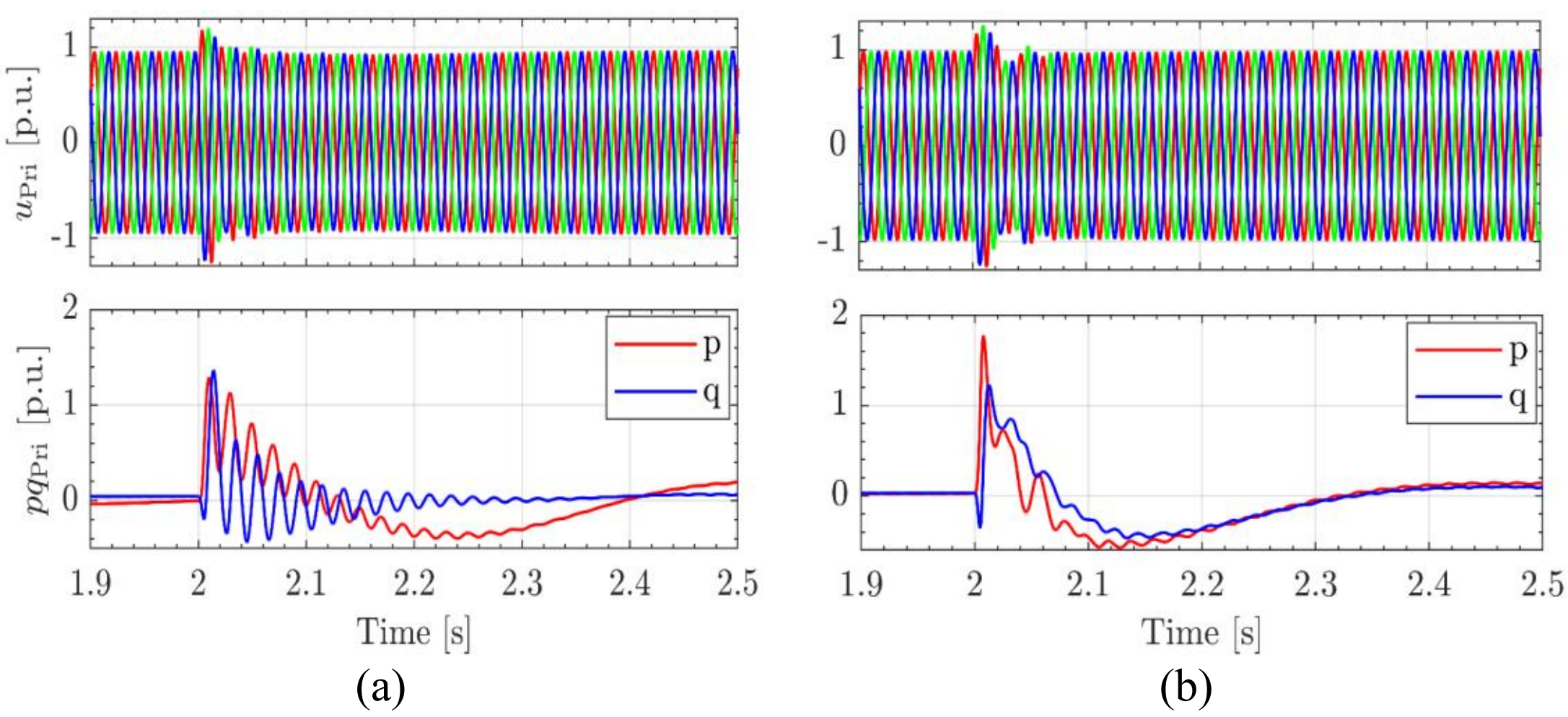

*Figure 5* – ***Results under 30⁰ phase jump with (a) Synchronous Condensers, (b) E-STATCOM***

## 3.2 Sub-Synchronous Oscillation (SSO)

The system in Fig. 4 is simulated under a three-phase to-ground fault by disconnecting the compensating devices. The fault occurs at 2 s and is cleared after 0.2 s. Upon fault recovery, the grid voltage ($u_{Grid}$) experiences a significant TOV of 2 p.u. and an SSO of 23 Hz, as shown in Fig. 6(a). Additionally, the grid power oscillates at 23 Hz. The Synchronous condenser reduces the TOV and dampens the SSO within 4-5 cycles, as illustrated in Fig. 6(b). However, it injects a substantial amount of DC through $i_{DRC}$ during the fault event, which can degrade the lifetime of the breakers and potentially lead to their failure. In contrast, the E-STATCOM is more effective, further limiting the TOV, eliminating the DC component, and damping the SSO within just one cycle.

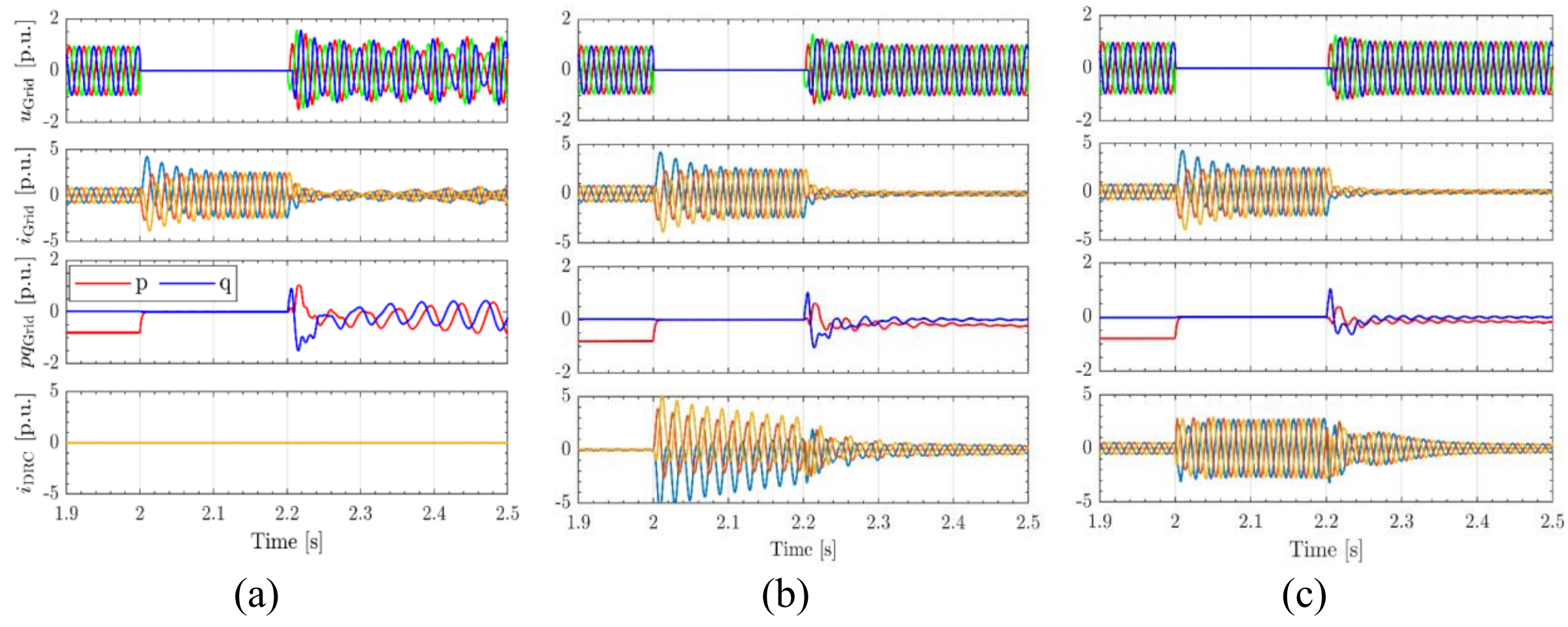

*Figure 6* – ***Results under three-phase to ground fault, (a) without compensation, (b) with Synchronous condenser, (c) with E-STATCOM***

## 3.3 Inertia and frequency control

A RoCoF of 2.5 Hz/s is initiated at 2 s, which continues until 2.4 s, as shown in Fig. 7. To ensure a fair comparison, the inertia constant (H) was set at 2 s for both the synchronous condenser and the E-STATCOM. Consequently, both devices inject the same peak power of 0.2 p.u. However, it is evident that the E-STATCOM not only injects power for a longer duration but also damps it faster once the system reaches a steady-state. Additionally, the performance of E-STATCOM can be easily adjusted based on requirements such as H value or damping factors. For instance, Fig. 7 demonstrates the ability of E-STATCOM to inject higher power with a higher H (= 4 s), a flexibility not achievable with the synchronous condenser. It is important to note that the H of a synchronous condenser is inherently tied to its physical characteristics, thereby limiting its achievable H values. This inherent limitation underscores the flexibility of the E-STATCOM, which can offer a wide range of adjustable H values to meet specific grid requirements.

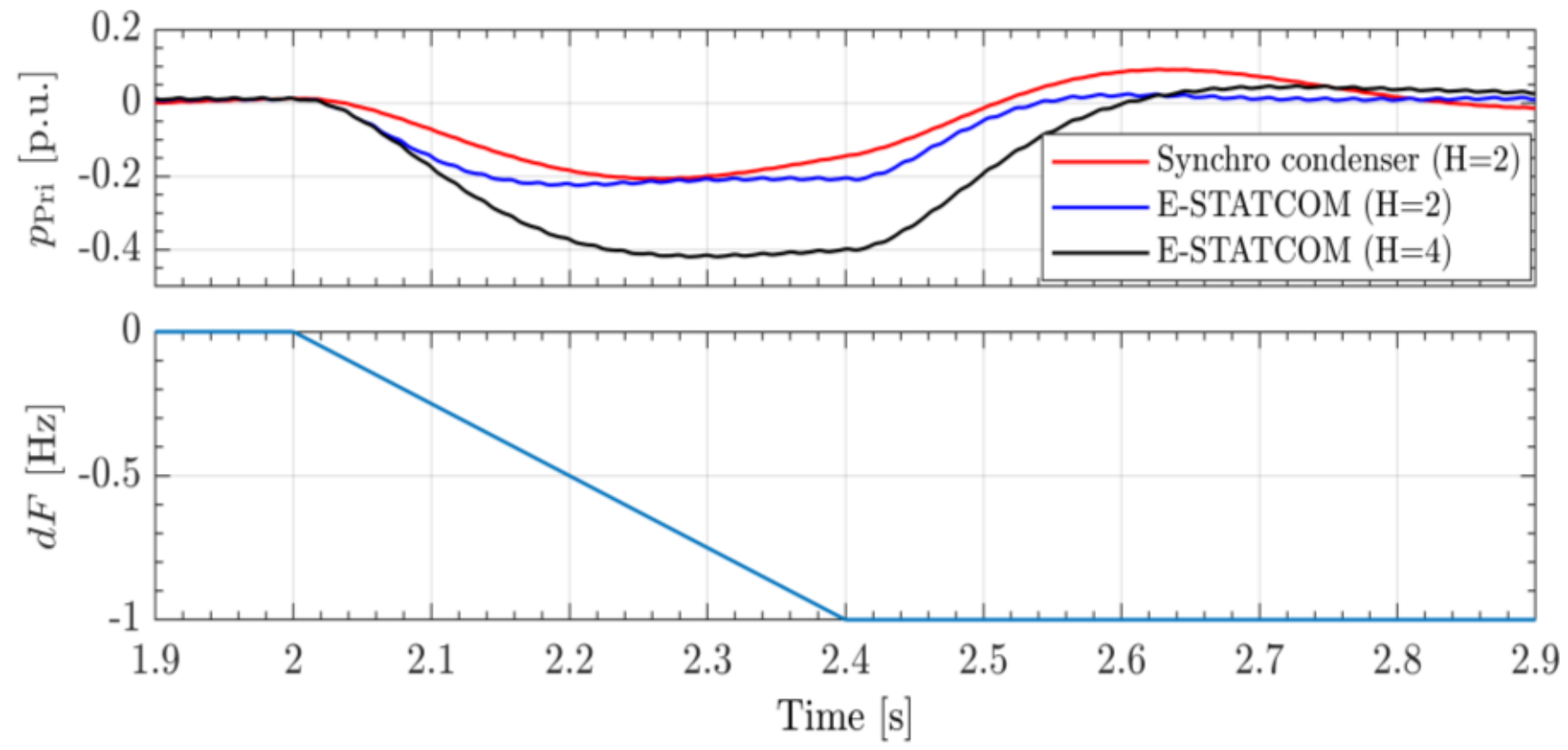

*Figure 7 – Power response to a RoCoF of 2.5Hz/s with Synchronous condenser and E-STATCOM.*

### 3.4 Fault current

A single-phase to ground fault is simulated for 0.2 s starting at 2 s, as shown in Fig. 8. It can be observed in Fig. 8(a) that the Synchronous condensers have no limitation on fault current injection, as it depends entirely on sub-transient reactance. In contrast, the E-STATCOM's fault current contribution is limited by its valve rating. However, an E-STATCOM with overcurrent capability can match the fault current contribution of the Synchronous condensers, as shown in Fig. 8(c). In such a case, the E-STATCOM can support full current during the recovery phase, unlike the Synchronous condensers.

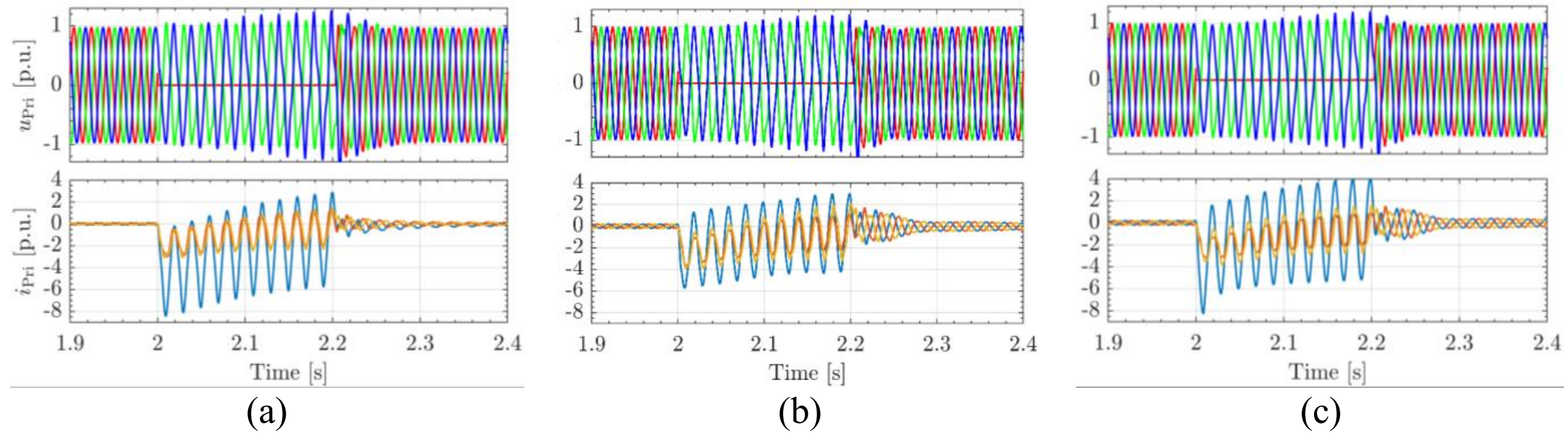

*Figure 8 –Results under Line-ground fault with (a) Synchronous condensers, (b) E-STATCOM, (c) E-STATCOM with overcurrent capability*

### 3.5 Losses

The loss evaluation is a significant consideration to operational cost for both STATCOM technology and Synchronous Condenser projects. The losses associated with STATCOM installations are significantly lower than those associated with synchronous condenser installations. Compared to a synchronous condenser, the full load losses of the STATCOM are about half the losses of the SCS. The maximum losses for the STATCOM is around 1% of the total rating. At no-load, the losses can be up to ten times less for the STATCOM than the SCS. The maximum no load losses for STATCOM are around 0.1% of the total rating. The energy storage is of low power loss design, i.e. low leakage current and low equivalent series resistance resulting in low losses. Even though the rating of the STATCOM is increased to meet the short-circuit needs of the system compared to SCS, the loss evaluation favors the STATCOM technology due the typically heavy weighting for no load where this type of technology shall operate most of the time [11]. This is specifically seen for projects with high cost for losses since the intent of the loss evaluation is to capture the cost of losses over the life of the project.

### 3.6 Footprint

The required footprint for STATCOM technology and Synchronous Condenser Solution is strongly dependent on the services required for the power system. Typically, the footprint is of the same size if the required service is to provide short circuit current up to the capability of STATCOM system with one single converter. For projects requiring a large amount of inertia, the required footprint will be significantly larger for Synchronous Condenser Solutions due to the inefficient change of energy with the power system, i.e. several machines required to do provide the same inertia contribution as the E-STATCOM. Maintenance access is another consideration that will have a direct impact on the footprint, and for synchronous condenser project footprint will be significantly larger if additional space is required for maintenance of rotor with lifting device.

## 4 Conclusion

Through a comparative time-domain as well as frequency-domain simulation analysis, this paper demonstrates the main performance distinctions between GFM STATCOM/E-STATCOM and Synchronous condensers in terms of damping DC components, sub-synchronous resonance mitigation, inertia capability, and fault current capabilities.

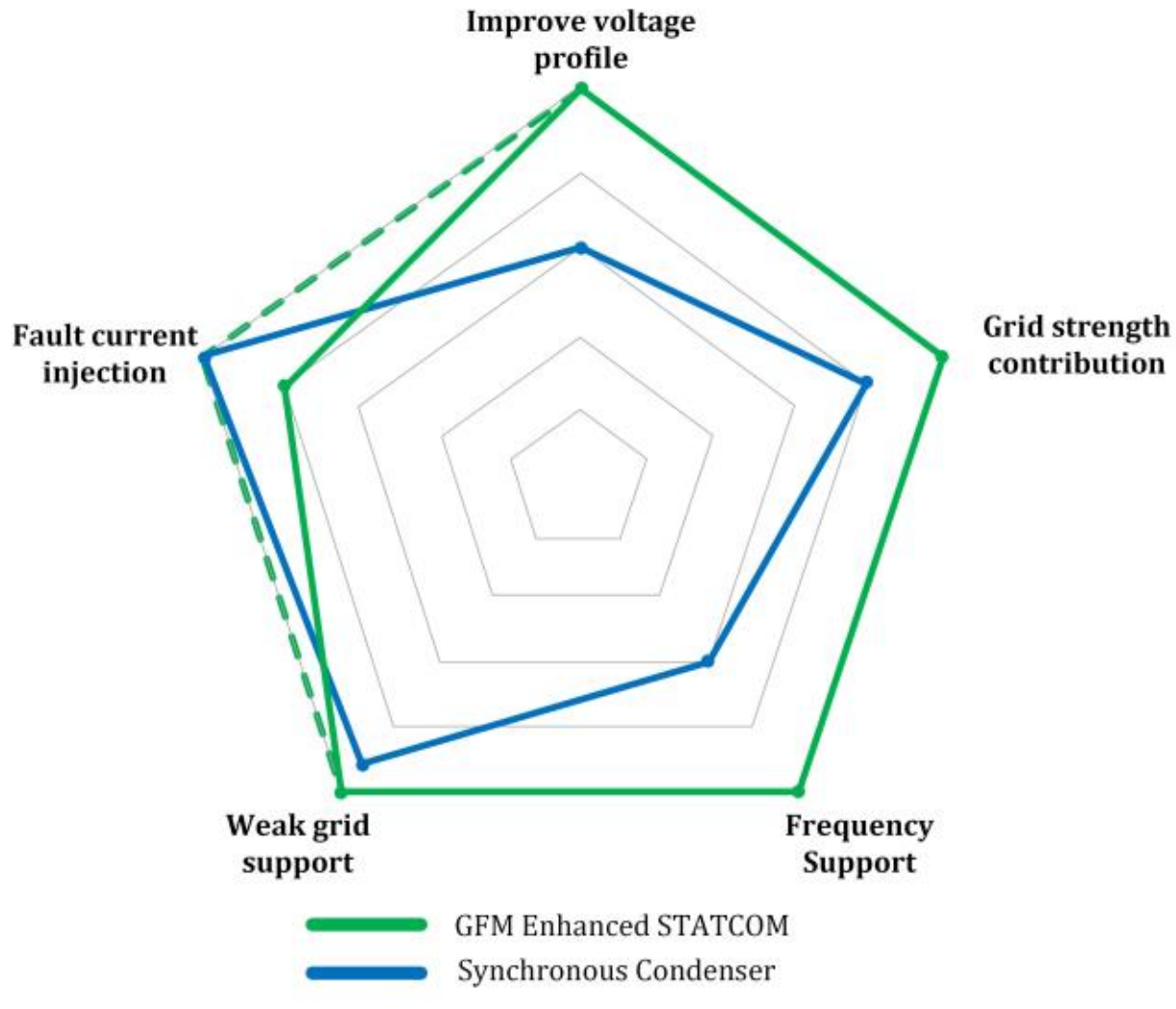

*Figure 9* ***- Summary of capabilities of technologies: GFM E-STATCOM, Synchronous condensers***

As depicted in Section 3, SIL is conducted to demonstrate the performance of grid-forming E-STATCOMs compared to synchronous condensers. The major differences between synchronous condensers and E-STATCOMs in managing grid stability are illustrated in the pentagon graph shown in Fig. 9. and outlined as follows:

1) **DC component damping:** Synchronous condensers inherently possess a high DC component during faults, whereas E-STATCOMs can be finely tuned to dampen these DC components effectively.
2) **Sub-synchronous resonance (SSR) and power oscillations:** Synchronous condensers are prone to SSR and power oscillations due to mechanical interactions and inertia, which can exacerbate instability in grids with high levels of inverter-based resources. In contrast, STATCOM and E-STATCOMs can be designed to avoid resonance, thereby offering better support for weak grids with high renewable penetration.
3) **Fault current capability:** While synchronous condensers have a higher fault current capability, appropriately rated STATCOM and E-STATCOMs can provide robust fault current support.
4) **Control tunability:** The parameters of STATCOM and E-STATCOMs can be dynamically tuned to optimize grid support, whereas synchronous condenser parameters are constrained by the machine's physical characteristics. This tunability of STATCOM and E-STATCOMs enables them to adapt to varying grid conditions and provides enhanced stability support for today's and tomorrow's fast-changing power grids.
5) **Footprint:** To achieve the same performance as an E-STATCOM for inertia contribution, multiple synchronous condensers may be required, which would significantly increase the overall footprint.
6) **Losses and maintenance:** STATCOM and E-STATCOMs typically have lower operational losses and maintenance requirements compared to synchronous condensers, contributing to their overall efficiency and reliability.